\documentclass[twocolumn, amsmath, aps, prl]{revtex4}
\usepackage{graphicx}
\usepackage{times}
\usepackage{epsfig}

\begin{document}
\title{Theoretical study of the accuracy limits for the optical resonance frequency measurements.}

\author{L. Labzowsky$^{1),2)}$, G. Schedrin$^{1)}$, D. Solovyev$^{1)}$ and
G. Plunien$^{3)}$}

\affiliation{ 1) V. A. Fock Institute of Physics, St. Petersburg
State University, Petrodvorets,
Oulianovskaya 1, 198504, St. Petersburg, Russia\\
2) Petersburg Nuclear Physics Institute, 188300, Gatchina, St.
Petersburg, Russia\\
3) Institut f\"{u}r Theoretische Physik Technische Universit\"{a}t
Dresden, Mommsenstra\ss e 13, D-01062, Dresden, Germany}

\begin{abstract}
The principal limits for the accuracy of the resonance frequency
measurements set by the asymmetry of the natural resonance line
shape are studied and applied to the recent accurate frequency
measurements in the two-photon 1s-2s resonance and in the
one-photon 1s-2p resonance in hydrogen atom. This limit for 1s-2s
resonance is found to be $\sim 10^{-5}$ Hz compared to the
accuracy achieved in experiment $\pm 46$ Hz. In case of deuterium
atom the limit is essentially larger: $10^{-2}$ Hz. For 1s-2p
resonance the accuracy limit is $0.17$ MHz while the uncertainty
of the recent frequency measurement is about $6$ MHz.
\end{abstract}
\maketitle

 The resonance approximation which reduces
the description of the spectral line profile to the Lorentz
contour is the cornerstone of all the modern transition frequency
measurements in the resonance experiments. This approximation is
employed for the extraction of the resonance frequency values from
the experimental data. Here we address only the natural line
profile, disregarding Doppler and collisional broadening. We
assume that the latter effects can be minimized below the level of
the natural line width. In the resonance approximation the line
profile is described by two parameters, resonance frequency
$\omega^{res}$ and the width $\Gamma$ and is symmetric with
respect to $\omega^{res}$. In the early paper by F. Low \cite{Low}
it was pointed out that the resonance approximation is valid only
up to a certain limit of accuracy which is defined by nonresonant
(NR) corrections. Beyond this limit the line profile acquires
asymmetric form and the definition of $\omega^{res}$ becomes
nonunique. Moreover, the NR corrections are, in principle,
process-dependent, so the frequency value also will differ for
different excitation processes. This means that the measurement of
frequency is a well defined procedure only up to some limit, which
we define in our paper. Going beyond this limit, it should be
replaced by the direct measurement of the spectral line profile,
which then becomes the only quantum-mechanical observable.

During the last few years the NR corrections were reconsidered in
a series of theoretical works \cite{Asimmetry}-\cite{PRA2002}.
This reconsideration was triggered by the new extra accurate
optical resonance experiments \cite{Eikema}-\cite{Niering}. In the
experiment \cite{Eikema} the natural line profile for the
Lyman-$\alpha$ 1s-2p one-photon resonance in hydrogen was observed
for the first time. In \cite{Huber}, \cite{Niering} the most
accurate measurements in the optical region were performed for the
two-photon 1s-2s resonance in hydrogen. In the latter experiments
the absolute accuracy of the frequency measurements was as high as
$\pm 46$ Hz, or $10^{-14}$ in relative units. However, the scheme
of these experiments causes serious difficulties for the
theoretical description of NR corrections and hence for the
definition of the accuracy limits in this case. In \cite{Huber},
\cite{Niering}, the H atoms were excited from the ground 1s state
to the 2s state via two-photon absorption in the space region free
of external fields. After this the excited atoms moved during
approximately $10^{-3}$ s to another space region where they meet
a weak electric field. In this field 2s  and 2p states are mixed
and the atoms decay via the ordinary 2p-1s transition. This
radiation is detected and provides the necessary information for
the extraction of 1s-2s frequency value from the experimental
data. In the rest frame of an atom it looks like the excitation
occurs in the absence of the electric field and then the electric
field is turned on (delayed decay). In terms of QED the initial
and final states of an atom are described then by different
Hamiltonians $\widehat{H}_{in}$ and $\widehat{H}_{out}$.

The QED theory with different in- and out-Hamiltonians was
developed by Fradkin, Gitman and Shvartsman (see book
\cite{Fradkin}). This theory we will use below for the description
of NR corrections in the two-photon 1s-2s resonance experiment.
The earlier attempts \cite{Jentshura}-\cite{Canadian} to describe
the NR corrections in these experiments with standard QED methods
cannot be considered as reliable.

The major contribution to NR corrections arises from the
interference between the resonant and nonresonant terms in the
expression for the scattering amplitude \cite{Asimmetry}. In
\cite{Asimmetry} only the nonresonant terms with the same symmetry
as the resonant one were included. As it was pointed out in
\cite{Jentshura} the most important contribution arises from the
interference between the resonant term and nonresonant terms with
another symmetry: the NR contribution of $2p_{3/2}$ state to
$1s-2p_{1/2}$ resonance was considered as an example. It was
argued that this contribution can survive in differential (with
respect to the angles) cross-section for the resonance photon
scattering on an atom. In the present work we will show that the
NR contributions of this type survive both in differential and
total cross-section provided that the resonance line shape is
fully natural. Moreover, in many cases the dominant NR
contributions which define the accuracy limits for the frequency
measurements arise from the neighbor hyperfine (HF) components.
These components have different symmetry (values of the total
angular momentum $F$ of an atom) compared to the HF component
which is used in the basic transition (see Fig. 1). The
contribution of HF neighbor components was missing in the earlier
investigations \cite{Asimmetry}-\cite{PRA2002}.

We start with the simplest case of the elastic photon scattering
on the hydrogen atom in the absence of the electric field. The
amplitude for this process can be written in a standard form
\cite{Akhiezer}: \onecolumngrid
\begin{eqnarray}
\label{1}
 A^{JM_f,JM_i}_{n_0j_0l_0m_f;n_0j_0l_0m_i}=\sum\limits_{njlm}\frac{<n_0j_0l_0m_f|A^*_{JLM_f}|njlm>
 <njlm|A_{JLM_i}|n_0j_0l_0m_i>}{E_{njl}-E_{n_0j_0l_0}-\omega}
\end{eqnarray}

\twocolumngrid

where the quantum numbers $njlm$ represent the standard set of
one-electron quantum numbers for the electron in the hydrogen
atom. The photons are characterized by the total angular momentum
$J$, its projection $M$ and the orbital angular momentum $L$. The
latter defines parity, or the type of the photon: electric or
magnetic. The operator $A_{JLM}$ corresponds to photon absorption,
$A^*_{JLM}$ denotes photon emission. In Eq. (\ref{1}) the
additional term where the operators $A^*$ and $A$ are interchanged
and the sign of $\omega$ is reversed, is omitted. This term
vanishes in the resonance approximation and does not contribute to
the leading NR corrections. The resonance condition for the
frequency is $\omega=E_{n_1j_1l_1}-E_{n_0j_0l_0}$ and in the
resonance approximation in Eq. (\ref{1}) only the terms  with
$njl=n_1j_1l_1$ are retained.

 To avoid the singularity in Eq. (\ref{1}) we insert
in the denominator the width of the excited state
$\Gamma_{n_1j_1l_1}$. The regular method for this insertion within
the framework of QED is described in \cite{Low}, \cite{Akhiezer},
\cite{LabKlim}, \cite{Labz_Proz}.

With this description of the resonant process the NR corrections
arise as the other terms of the expansion in Eq. (\ref{1}) with
$njl\neq n_1j_1l_1$. Performing the summations over all the
angular momentum projections, we obtain for the process
probability in the resonance approximation:
\begin{eqnarray}
\label{4}
 W_{n_0j_0l_0,n_0j_0l_0}=\frac{S^{1\gamma}(j_0j_1j_1J)}
 {(E_{n_1j_1l_1}-E_{n_0j_0l_0}-\omega)^2+\frac{1}{4}\Gamma^2_{n_1j_1l_1}}\times
 \\
 \nonumber
 \frac{1}{2\pi}W^{em}_{n_0j_0l_0,n_1j_1l_1}W^{ab}_{n_1j_1l_1, n_0j_0l_0}
\end{eqnarray}
where $W^{ab}$, $W^{em}$ are the standard absorption and emission
probabilities per time unit and
$S^{1\gamma}(j_0j_1j_2J)=\sum\limits_a\left\{
\begin{array}{ccc}
  j_0 & j_1 & J \\
  J & J & a \\
  j_2 & j_0 & J
\end{array}\right\}$. Taking into account the NR
correction due to the closest to $n_1j_1l_1$ neighbor level
$n_2j_2l_2$ results in \onecolumngrid
\begin{eqnarray}
\label{5}
W_{n_0j_0l_0,n_0j_0l_0}=\frac{1}{2\pi}\left[\frac{W^{em}_{n_0j_0l_0,n_1j_1l_1}W^{ab}_{n_1j_1l_1,
n_0j_0l_0}}
 {(E_{n_1j_1l_1}-E_{n_0j_0l_0}-\omega)^2+\frac{1}{4}\Gamma^2_{n_1j_1l_1}}S^{1\gamma}(j_0j_1j_1J)+\right.
 \\
\left. \nonumber
 +2Re\frac{A^{em*}_{n_0j_0l_0,n_1j_1l_1}A^{em}_{n_0j_0l_0,n_2j_2l_2}A^{ab*}_{n_1j_1l_1,n_0j_0l_0}A^{ab}_{n_2j_2l_2,n_0j_0l_0}}{(E_{n_1j_1l_1}-E_{n_0j_0l_0}-\omega-\frac{i}{2}\Gamma_{n_1j_1l_1})
 (E_{n_2j_2l_2}-E_{n_0j_0l_0}-\omega-\frac{i}{2}\Gamma_{n_2j_2l_2})}S^{1\gamma}(j_0j_1j_2J)
\right]
\end{eqnarray}\twocolumngrid
Here $A^{em}$, $A^{ab}$ are the reduced emission and absorption
amplitudes (reduced matrix elements of the photon emission and
absorption operators). These matrix elements do not depend on the
angular momentum projections. The factor $S^{1\gamma}(j_0j_1j_2J)$
does not vanish for $j_2\neq j_1$, so that the dominant NR
corrections due to the interference of the resonant and
nonresonant terms with different symmetry ($j_1l_1\neq jl$) can
exist. They exist both in differential (with respect to the
angles) and total probability; Eq. (\ref{5}) describes the total
probability. The differential one would depend on the angle
$(\vec{n_i}, \vec{n_f})$ where $\vec{n_i}$, $\vec{n_f}$ are the
velocity directions for the absorbed and emitted photons. The
vector $\vec{n_i}$ is fixed by the incident laser beam and
$\vec{n_f}$ by the position of the detector.

Now we turn to the case of the 1s-2s two-photon absorption and
delayed decay in electric field. The formalism of QED with
different in- and out-Hamiltonians developed in \cite{Fradkin}
pursues the more complicated task: the creation of the
electron-positron pairs in the strong electric field. We will
apply this formalism to the case of a weak electric field.
However, the situation in our case is also non-perturbative: due
to the admixture of the $2p$ state into the $2s$ state the
emission probability changes by 8 orders of magnitude. The change
of the other atomic characteristics (Stark shift of the energy
levels, Stark splitting) we will consider as negligible. The
criterion of the weak field will be $\varepsilon<
\varepsilon_c=475$ V/cm, where $\varepsilon$ is the strength of
the electric field. In the field $\varepsilon=\varepsilon_c$ the
2s and 2p levels are 100$\%$ mixed \cite{Bethe}.

The Fradkin-Gitman-Shvartsman (FGS) theory \cite{Fradkin} operates
with two complete sets of eigenfunctions belonging to in- and
out-Hamiltonians. This theory follows the standard QED approach in
generalized form: S-matrix, field operators in the Fock space,
4-dimensional perturbation expansion for the S-matrix elements,
Wick theorem and Feynman graph techniques. Actually  the unique
new element that we will have to use is the generalized FGS
electron propagator. This propagator connects two vertices, which
are described by in- and out-Hamiltonians, respectively.

In our case this means the absence or presence of the electric
field. The FGS propagator looks like
\begin{eqnarray}
\label{6}\
 S^{FGS}(x_1x_2)=\Theta(t_1-t_2)\sum\limits_{
\tilde{n}, n(E_{\tilde{n};n}>0)
 }\psi_{\tilde{n}}(x_1)\omega_{\tilde{n}n}\bar{\psi}_n(x_2)
\\
\nonumber
 -\Theta(t_2-t_1)\sum\limits_{\tilde{n},
n(E_{\tilde{n};n}<0
)}\psi_n(x_1)\omega_{n\tilde{n}}\bar{\psi}_{\tilde{n}}(x_2)
\end{eqnarray}
where $\psi_{\tilde{n}}(x)$ are the solutions of the Dirac
equation for the electron in the field of the nucleus and the
external electric field, $\psi_{n}(x)$ are the solutions with zero
external field; $E_{\tilde{n}}$, $E_n$ are the corresponding
eigenvalues. The matrix $\omega_{\tilde{n}n}$ in the weak-field
limit reduces to an overlap integral
\begin{eqnarray}
\label{7}
 \omega_{\tilde{n}n}=\int\psi^+_{\tilde{n}}(\vec{x})\psi_n(\vec{x})d\vec{x}\equiv
 <\tilde{n}|n>
\end{eqnarray}
In the nonrelativistic limit, evidently valid for the neutral
hydrogen atom, we replace the Dirac wave functions by
Schr\"{o}dinger ones and omit the negative-energy contribution in
Eq. (\ref{6}).

Within the FGS theory, the probability of absorption of two
equivalent laser photons with frequency $\omega'$ by the electron
in the hydrogen atom in its ground state with the subsequent
delayed decay in external electric field in the resonance
approximation ($n=a$, $\tilde{n}=\tilde{a'}$) looks like
\begin{eqnarray}
\label{12}
 dW_{\tilde{a}a}=\frac{1}{2\pi}\frac{W^{(em)}_{\tilde{a}\tilde{a'}}|<\tilde{a'}|a'>|^2W^{(ab\, 2\gamma)}_{a'a}}
 {(E_{a'}-E_a-2\omega')^2+\frac{1}{4}\Gamma^2_{a'}}S^{2\gamma}_{aa'}
\end{eqnarray}
where $W^{(ab 2\gamma)}$ is the two-photon absorption probability,
$S^{2\gamma}_{aa'}$ is the angular factor similar to $S^{1\gamma}$
in Eq. (\ref{4}), and the resonance condition is
$E_{a'}-E_a=2\omega'$.

Taking into account the NR correction due to the closest to $a'$
neighbor level $a''$ results in an expression similar to Eq.
(\ref{5})\onecolumngrid
\begin{eqnarray}
\label{13}
 dW_{\tilde{a}a}=\frac{1}{2\pi}\left\{\frac{W_{\tilde{a}\tilde{a'}}^{em}|<\tilde{a'}|a'>|^2W^{(ab\, 2\gamma)}_{a'a}}
 {(E_{a'}-E_a-2\omega')^2+\frac{1}{4}\Gamma_{a'}^2}S^{2\gamma}_{aa'}+
\right.
\\
\nonumber
 \left.
 +2Re\left[\frac{A^{em*}_{\tilde{a}\tilde{a'}}<\tilde{a'}|a'>^*A^{em}_{\tilde{a}\tilde{a''}}<\tilde{a''}|a''>
 A^{(ab\, 2\gamma)*}_{a'a}A^{(ab\, 2\gamma)}_{a''a}}
 {(E_{a'}-E_{a}-2\omega'-\frac{i}{2}\Gamma_{a'})(E_{a''}-E_a-2\omega'-\frac{i}{2}\Gamma_{a''})}\right]S^{2\gamma}_{aa''}
 \right\}
\end{eqnarray}
\twocolumngrid where $A^{(ab\, 2\gamma)}_{a'a}$, $A^{(ab\,
2\gamma)}_{a''a}$ are the reduced two-photon absorption
amplitudes.

 For moving further we have to choose the procedure
for determination of the resonance photon frequency. In
\cite{Asimmetry}-\cite{PRA2002} the evaluation of the maximum
value of the frequency distribution was used for this purpose. As
it was shown in \cite{Jentshura} any other procedure (e.g. finding
a "center of gravity" for the line profile) would give the result
quite close to the choice formulated above. In case of the Lorentz
profile all the methods of defining $\omega^{res}$ give the same
result $\omega_0^{res}=E_{a'}-E_a$. With our choice the NR
correction will look like
\begin{eqnarray}
\label{14}
 \delta\omega^{NR}=\omega^{max}-\omega^{res}_0
\end{eqnarray}
where $\omega^{max}$ is the frequency value, corresponding to the
maximum of the frequency distribution.

The first example, that we will consider, is the two-photon
$1s_{1/2}(F=1)+2\gamma\rightarrow 2s_{1/2}(F=1)$ transition in
hydrogen \cite{Huber}, \cite{Niering}. The value $K=1$ for the
total angular momentum $K$ of the two-photon system with equal
photon frequencies is strictly forbidden \cite{Akhiezer}. This
concerns exactly the case of the two-photon absorption of laser
photons in experiment \cite{Huber}, \cite{Niering} (see Fig. 1).
According to this picture, the main NR contribution arises from
the transition $1s_{1/2}(F=1)\rightarrow 2p_{1/2}(F=1)$. For
deriving the NR correction we have to use an expression (\ref{13})
where we have to set $a=1s_{1/2}(F=1)$, $a'=2s_{1/2}(F-1)$ and
$a''=2p_{1/2}(F=1)$. In a weak electric field
$\psi_{\tilde{a}}\simeq\psi_a$,
$\psi_{\tilde{a'}}\simeq\psi_{a'}+\eta\psi_{a''}$,
$\psi_{a''}\simeq\psi_{a''}-\eta\psi_{a'}$ and the overlap
integrals are $<\tilde{a}|a>\simeq<\tilde{a''}|a''>\simeq 1$. Here
$\eta=|\Delta E_S|/\Delta E_L$ is the Stark shift to the Lamb
shift ratio. This ratio is $\eta=1$ for a field
$\varepsilon=\varepsilon_c$. For deriving the NR correction we set
in Eq. (\ref{13}) $E_{a'}-E_a=\omega^{res}_0$,
$E_{a''}-E_{a}=\omega_0^{res}-\Delta E_L$, where $\Delta
E_L=E_{a''}-E_{a'}\approx 10^3$ MHz is the Lamb shift. The width
$\Gamma_{a'}$ in case of the experiment \cite{Huber},
\cite{Niering} is determined by the experimental set up (the time
delay before the excited atoms enter the electric field region)
and is equal to $\Gamma_{a'}=\Gamma_{exp}\approx 1$ kHz. Insertion
of the wave functions $\psi_{\tilde{a}}$, $\psi_{\tilde{a'}}$ and
$\psi_{\tilde{a''}}$ in the emission amplitudes yields:
$A^{em}_{\tilde{a}\tilde{a'}}\approx\eta(\Gamma_{2p})^{1/2}$,
$A^{em}_{\tilde{a}\tilde{a''}}\approx(\Gamma_{2p})^{1/2}$, where
$\Gamma_{2p}$ is the width of 2p state. Then, evaluating the
maximum value of Eq. (\ref{13}) with respect to $\omega'$ and
defining the NR correction according to Eq. (\ref{14}) we find
\begin{eqnarray}
\label{15}
 |\delta\omega^{NR}|\approx\frac{1}{4}\frac{\Gamma^2_{exp}}{\Delta
 E_L}\left(\frac{W^{2\gamma}_{1s,2p_{1/2}}}{W_{1s,2s}^{2\gamma}}\right)^{1/2}\frac{1}{\eta}\frac{S^{2\gamma}_{aa''}(F,F'')}
 {S^{2\gamma}_{aa'}(F,F')}
\end{eqnarray}
where $S^{2\gamma}_{aa'}(F,F')$ is the angular factor defined for
the transition between the two hyperfine sublevels. In the present
case $F=F'=F''=1$, and the corresponding angular factors are
equal: $S^{2\gamma}_{aa'}(F,F')=S^{2\gamma}_{aa''}(F,F'')=11/18$.
The probability for the two-photon transition $W^{2\gamma(E1E1)}
_{1s,2s}$ is very well known. An accurate nonrelativistic value
for this transition was obtained in \cite{Klarsfeld}:
$W^{2\gamma(E1E1)}_{1s,2s}=1.32\cdot 10^{-3}(\alpha Z)^6$ a.u.
Here $\alpha$ is the fine structure constant, $Z$ is the nuclear
charge and it is assumed that $\alpha Z\ll 1$. The probability
$W^{2\gamma}_{1s, 2p_{1/2}}$ was evaluated recently numerically
for all $Z$ values ($1\ll Z\ll 100$) \cite{Lab_Shon} and
analytically for $\alpha Z\ll 1$ \cite{EurPhys}. The result is
$W^{2\gamma}_{1s, 2p_{1/2}}=W^{2\gamma(E1M1)}_{1s,
2p_{1/2}}+W^{2\gamma (E1E2)}_{1s, 2p_{1/2}}=2.907\cdot
10^{-5}(\alpha Z)^8+3.69\cdot 10^{-6}(\alpha Z)^8$ (a.u.).

The dependence $\eta^{-1}$ in Eq. (\ref{15}) cannot be continued
to zero field; the meaningful limit is set by the field where the
decay rate of $2s$ level due to the admixture of $2p$ state becomes
equal to the natural decay width of $2s$ state.
This limiting field strength would be so small,
that it cannot be used in a real experiment. Inserting all the
numbers in Eq. (\ref{15}) and taking $\eta=0.1$ which corresponds
to the weak electric field $\varepsilon=47,5$ V/cm, we obtain the
final result $|\delta\omega^{NR}|\approx 10^{-5}$ Hz. This
accuracy limit is still far from the recent inaccuracy estimate in
the experiments \cite{Huber}, \cite{Niering}: $\pm 46$ Hz.
However, it is important to notice that the situation in hydrogen
seems to be rather fortunate for the accurate resonance frequency
measurement due to the absence of the transition to another
($F=0$) hyperfine sublevel of 2s state (see Fig. 1). For
comparison, in the deuterium where the total atom angular momentum
values for the $1s$ and $2s$ levels are $F=1/2$, $3/2$
respectively, this transition is allowed and NR correction by
order of magnitude is
\begin{eqnarray}
\label{16}
 |\delta\omega^{NR}|\sim\frac{1}{4}\frac{\Gamma^2_{exp}}{\Delta E_{HFS}}
\end{eqnarray}
where $\Delta
E_{HFS}=E_{2s_{1/2}(F=3/2)}-E_{2s_{1/2}(F=1/2)}\approx 100$ MHz is
the hyperfine-structure interval for 2s level. Taking the same
$\Gamma_{exp}$ value as in \cite{Huber}, \cite{Niering} we would
have $|\delta\omega^{NR}|\sim 10^{-2}$ Hz which is 3 orders of
magnitude larger than for hydrogen. This is not so far from the
accuracy limit of about 0.1 Hz that was considered in
\cite{Niering} as achievable in the future with the use of colder
hydrogen atoms.

A quite different situation arises for the 1s-2p resonant experiment
 in hydrogen \cite{Eikema}. In \cite{Eikema} the
resonance $1s(F=1)\rightarrow 2p_{3/2}(F=2,1)$ was measured. The
hyperfine structure for the $2p_{3/2}$ level was not resolved since
$\Gamma_{2p}>\Delta E_{HFS}(2p_3/2)$: $\Gamma_{2p}\approx 100$
MHz, $\Delta E_{HFS}(2p_{3/2})=E(2p_{3/2}, F=2)-E(2p_{3/2},
F=1)=23.7$ MHz. This is a typical case of overlapping
resonances for two hyperfine sublevels. However, due to the
presence of interference terms (see Eq. (\ref{5})) the line shape
deviates from the overlap of two Lorentz profiles and can be
presented by an expression:
\begin{eqnarray}
\label{17}
 F(\omega)=\frac{f(F,F')}{(\omega_1^{res}-\omega)^2+\frac{1}{4}\Gamma^2_{2p}}+
 \frac{f(F,F'')}{(\omega_2^{res}-\omega)^2+\frac{1}{4}\Gamma^2_{2p}}+
 \\
 \nonumber
 +2Re\frac{g(F,F',F'')}{(\omega_1^{res}-\omega-\frac{i}{2}\Gamma_{2p})(\omega^{res}_2-\omega-\frac{i}{2}\Gamma_{2p})}
\end{eqnarray}
where $\omega_1^{res}=E(2p_{3/2}, F=2)-E(1s_{1/2},F=1)$,
$\omega_2^{res}=E(2p_{3/2},F=1)-E(1s_{1/2, F=1})$, $f$, $g$ are
the angular factors, similar to $S^{1\gamma}$ in Eq. (\ref{4}).
These angular factors play an important role in defining the line
shape Eq. (\ref{17}): $f(1,2):f(1,1):g(1,2,1)=181:1:0.307$. Thus
the line shape exhibits the one-peak structure (the larger peak
fully screens the smaller one). The deviation from the overlap of
two Lorentz profiles is of order $0.307/181\approx0.17\%$. Then
the absolute accuracy limit for determination of
$1s(F=1)\rightarrow 2p_{3/2}$ transition frequency can be
estimated as $\Gamma_{2p}\cdot 0.0017\approx 0.17$ MHz. The
existence of the interference terms that distort the closely lying
resonances is well known (see for example, \cite{Baird}). However,
in \cite{Baird} this distortion was included in the error bars for
the transition frequencies not distinguishing between other contributions
of purely technical origin,
such as the laser intensity distribution and the spread of
atomic velocities. Nowadays, when the two latter effects in practice
can be diminished completely (e.g. by the use of colder atoms) it
becomes more useful to introduce such quantity as an "absolute limit"
for the frequency determination in the resonance experiments.
Actually any resonance experiment pretending on the utmost
accuracy should contain fittings of the experimental data with a
suitable theoretical line shape including the effects of
interference, originating from NR corrections. In case of
$1s-2p_{3/2}$ transitions Eq. (11) provides
such a theoretical line shape, derived from
first principles of QED.

\begin{center}
Acknowledgments
\end{center}
The authors are grateful to V. M. Shabaev, G. W. F. Drake, S. G.
Karshenboim and N. D. Filonov for valuable comments. L. L., D. S.
and G. P. acknowledge financial support from DFG, BMBF, GSI
and INTAS-GSI grant Nr. 0354-3604.
The work of L. L., G. S. and D. S. was also
supported by the RFBR grant Nr. 05-02-17483.

%\newpage
\begin{figure}
\mbox{\epsfxsize=5.5cm \epsffile{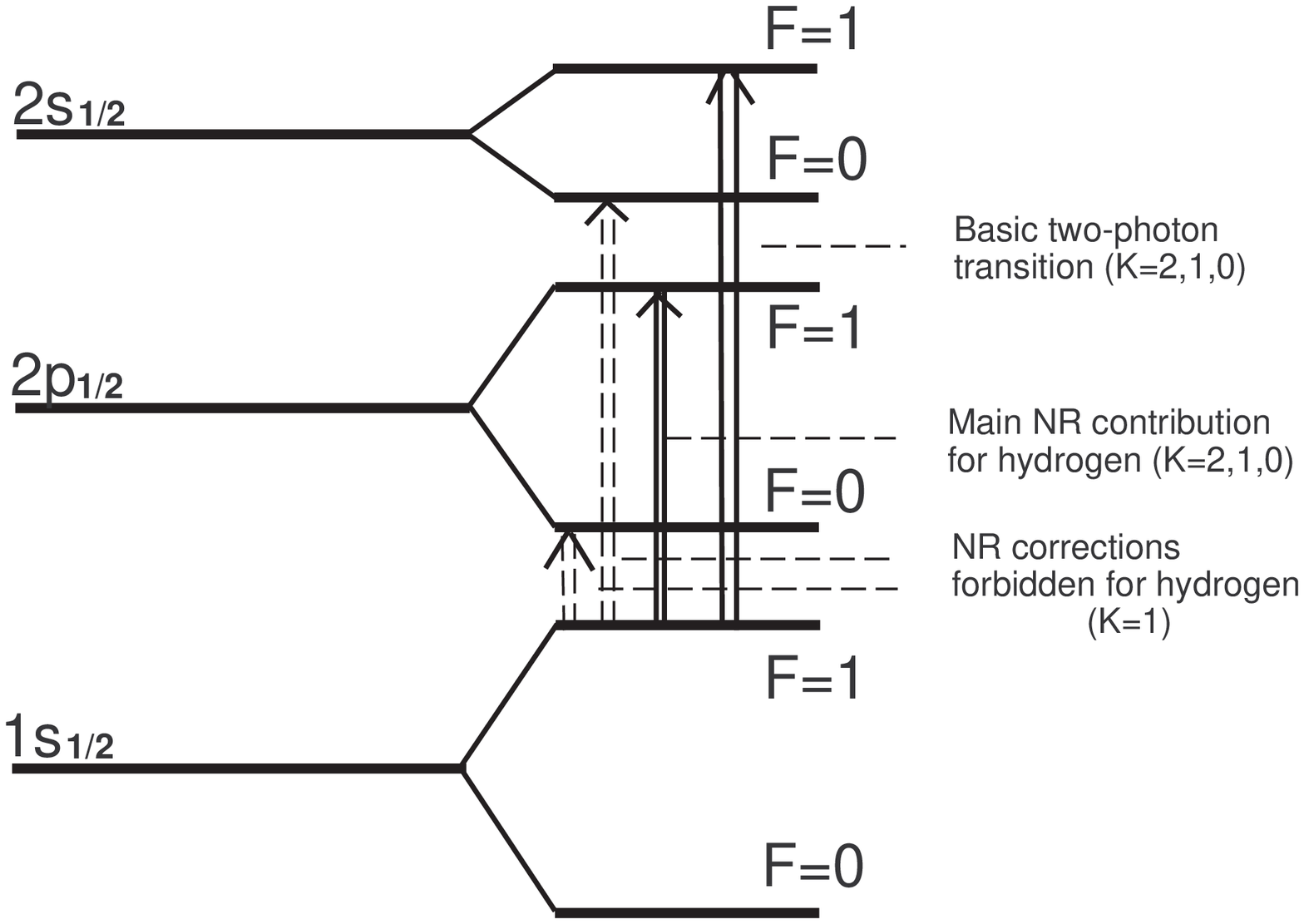}}

{\scriptsize Fig. 1. Scheme of the levels for the two-photon 1s-2s
transition. The vertical double lines denote the two-photon
transitions. The $K$ numbers denote the total angular momentum for
a two-photon system, possible for different two-photon
transitions.}
\end{figure}

\end{document}